\documentclass[aps, prl, showpacs, twocolumn,superscriptaddress]{revtex4}

\usepackage{graphicx, bm, amsmath, amsfonts,amssymb}

\usepackage{subfigure}

\begin{document}

\title{Nearly-flat bands with nontrivial topology}
\author{Kai Sun}
\affiliation{
Condensed Matter Theory Center and Joint Quantum Institute,
Department of Physics, University of Maryland, College Park, MD 20742, USA\\
}

\author{Zhengcheng Gu}
\affiliation{ Kavli Institute for Theoretical Physics, University of
California, Santa Barbara, CA 93106, USA\\
}

\author{Hosho Katsura}
\affiliation{
Department of Physics, Gakushuin University, Mejiro, Toshima-ku, Tokyo 171-8588, Japan\\
}

\author{S. Das Sarma}
\affiliation{
Condensed Matter Theory Center and Joint Quantum Institute,
Department of Physics, University of Maryland, College Park, MD 20742, USA\\
}

\date{\today}
\begin{abstract}
We report the theoretical discovery of a class of 2D tight-binding models containing nearly-flat bands with nonzero Chern numbers. In contrast with previous studies, where nonlocal hoppings are usually required, the Hamiltonians of our models only require short-range hopping 
and have the potential to be realized in cold atomic gases. Due to the similarity with 2D continuum Landau levels, these topologically nontrivial nearly-flat bands may lead to the realization of fractional anomalous quantum Hall states and fractional topological insulators in real materials. Among the models we discover, the most interesting and practical one is a square-lattice three-band model which has only nearest-neighbor hopping. To understand better the physics underlying the topological flat band aspects, we also present the studies of a minimal two-band  model on the checkerboard lattice.
\end{abstract}


\maketitle

In fermionic systems, 
a flat band (a macroscopically degenerate manifold of single-particle states) plays an important role in the study of strongly correlated phenomena, due to the vanishing bandwidth. One way to achieve a flat band is via the destructive interference of electron hoppings, which gives rise to moderately localized single-particle eigenstates~\cite{Weaire-Thorpe, Straley,Kohmoto1, Mielke, Tasaki,
Schulenburg, Tsunetsugu,C_Wu,Wu}. 
Landau levels, which are formed when a magnetic field is applied to a 2D electron gas (2DEG), can be considered as another type of flat bands arising in continuum rather than lattice 2D systems.
Different from the examples above, a Landau level has a nontrivial topological index (the Chern number). When an integer (or certain fractional) number of Landau levels is filled,
the system turns into an insulator with nontrivial topology,
known as the integer (or fractional) quantum Hall effect (IQHE or FQHE)~\cite{Stormer1999}.

In addition to 2DEG, efforts have also been made to realize IQH and FQH effects in lattice systems without magnetic field.
The first and most celebrated example is the anomalous quantum Hall state proposed by
Haldane~\cite{Haldane1988}. More recently, a new class of topological states, the time-reversal invariant topological insulators
characterized by a $\mathbb{Z}_2$ topological index is discovered in various lattice systems 
(see the recent reviews of Refs. \cite{Hasan2010,Qi2010} and
references therein). These lattice topological states share strong similarities with the IQH states.
However, the lattice counterpart of the fractional quantum Hall
states has not yet been discovered. One of the key
challenges to reach a fractional topological state in lattice models lies in the fact that
the bandwidth of a topologically nontrivial band in these models is
usually comparable or even larger than the band gap. Thus, at
fractional filling, the system is expected to be a Fermi liquid while interaction effects are
just subleading corrections. Therefore, a flat band with nontrivial topology is expected
to be the key in realizing lattice fractional topological states similar to the FQHE.
Recently, there have been some attempts to find completely flat bands with nonzero Chern numbers
in 2D lattice models \cite{Katsura1, Green}.
However, it turned out that, in all examples  except in the quasi-one dimensional thin torus~\cite{Katsura1},
flat bands have zero Chern number.

Since a topological index remains invariant under adiabatic
deformations as long as the gap is preserved, a straightforward
way to form such a flat band is to use the {\it spectral flattening trick}, i.e., an adiabatic
transformation from the original Hamiltonian to a new one with completely flat bands.
This technique is used in the classification of topological insulators and superconductors
~\cite{Qi2008, Ryu, Kitaev2009}. However, 
such a procedure 
may results in 
long-range hopping making the Hamiltonian nonlocal (see Supplementary Material for details).

In real materials, the exact flatness for a band is not a
physical requirement and we can relax
the constraint a little bit by allowing the band to have a nonzero bandwidth but requiring the
bandwidth to remain much smaller than the band gap. Unfortunately,
to the best of our knowledge, even such models have never
been reported. In this letter, we propose a generic
scheme to produce such kind of models based on a special class of
tight-binding Hamiltonians with short-range hoppings.
The band structure of these models contains nontrivial band touchings
 with quadratic dispersions (in contrast to the linear ones
near a Dirac point),
which are protected by the
time-reversal and lattice point-group symmetries as well as the
nontrivial topology \cite{sun2009, sun2010}. When the time-reversal
symmetry is broken, a band gap opens up at the band touching point
and the bands can have nonzero
Chern numbers. By slightly tuning
the short-range hopping strength, we find that some of the topologically
nontrivial bands can become nearly-flat.
We believe that this mechanism is very general and applies to
any tight-binding model with quadratic band touchings.
Surprisingly, in some of these models, this nearly-flat band situation is found
even with only nearest-neighbor (NN) hopping. These
nearly-flat bands have a strong analogy to
the Landau levels in 2DEG and thus may set the stage for exploring
new fractional topological states.

{\it Topological flat band with extremely short-range hopping: the square-lattice model
}--- Consider a square lattice with two space-inversion odd and one space-inversion
even orbitals per site, e.g. the $p_x$, $p_y$ and $d_{x^2-y^2}$ orbitals. This
model has been demonstrated in optical lattice systems~\cite{sun2010}.
In $k$-space, the Hamiltonian is
\begin{widetext}
\begin{align}
&H
=\sum_{\vec{k}}
\left(
\begin{array}{c c c}
d_{\vec{k}}^\dagger, & p_{x,\vec{k}}^\dagger, & p_{y,\vec{k}}^\dagger
\end{array}
 \right)
\left(
\begin{array}{c c c}
-2 t_{dd} (\cos k_x+\cos k_y)+\delta & 2 i t_{pd} \sin k_x& 2 i t_{pd} \sin k_y \\
-2 i t_{pd} \sin k_x &2 t_{pp} \cos k_x-2 t'_{pp} \cos k_y & i \Delta\\
-2 i t_{pd} \sin k_y &-i \Delta & 2 t_{pp} \cos k_y-2 t'_{pp} \cos k_x  \\
\end{array}
\right)
\left(
\begin{array}{c}
d_{\vec{k}} \\
p_{x,\vec{k}}\\
p_{y,\vec{k}} \\
\end{array}
\right),\nonumber
\end{align}
\end{widetext}
where $d_{\vec{k}}$, $p_{x,\vec{k}}$ and $p_{y,\vec{k}}$ are the fermion
annihilation operators at momentum $\vec{k}$ and the
lattice constant is set to unity.
The NN hoppings between
various orbitals are described by the hopping amplitudes
$t_{dd}$, $t_{pd}$, $t_{pp}$ and $t'_{pp}$. The $\delta$ term measures the
energy difference between $p$ and $d$ orbitals. The term ($\Delta$)
breaks the symmetry between the states with angular momentum $\pm1$
($p_x\pm i p_y$). This term breaks the time-reversal symmetry and
allows the Chern number to take a nontrivial value.

\begin{figure}
\begin{center}
\vskip 0.0cm \hspace*{-0.0cm}
\subfigure[]{\includegraphics[width=0.15\textwidth]{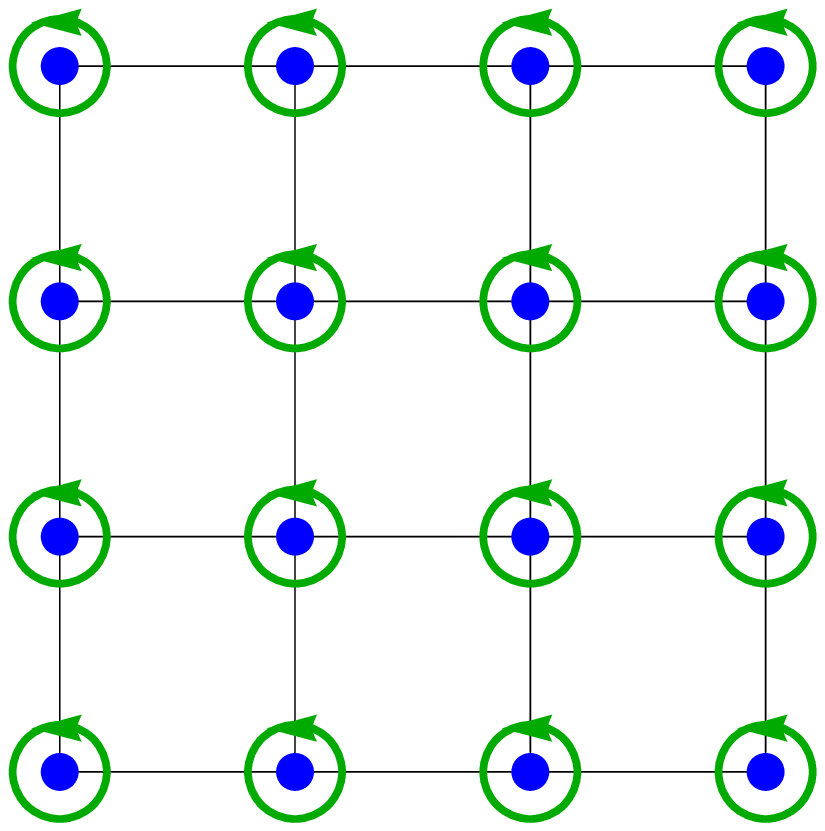}\label{fig:square_lattice}}
\subfigure[]{\includegraphics[width=0.25\textwidth]{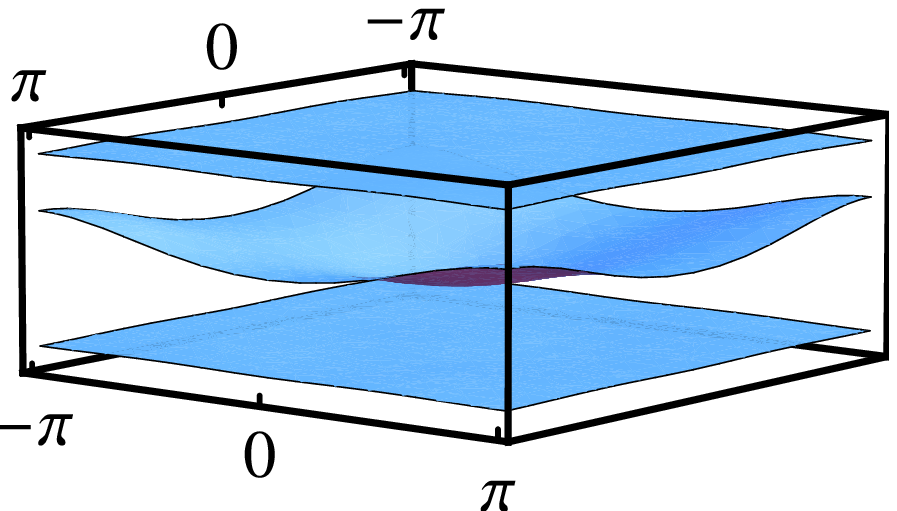}\label{fig:square_band}}
\subfigure[]{\includegraphics[width=0.4\textwidth]{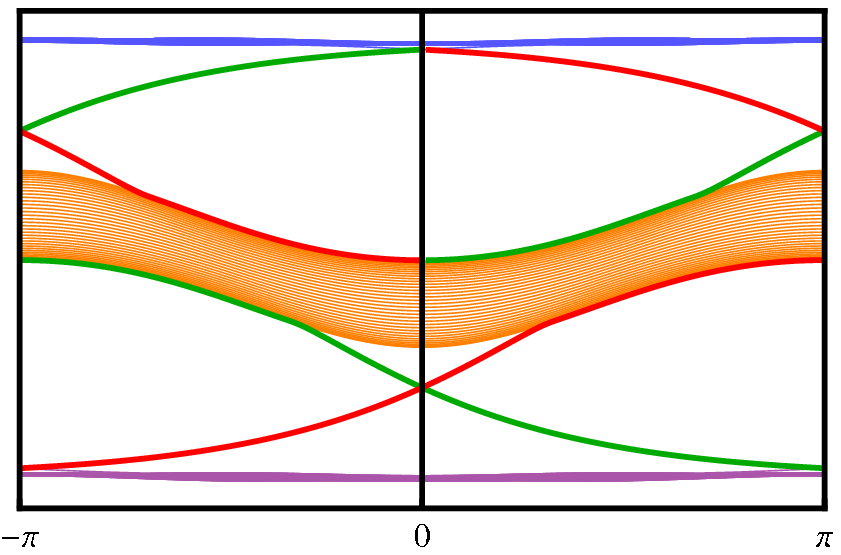}\label{fig:square_edge}}
\end{center}
\vskip-0.5cm \caption{(Color online) Chiral quasi-flat band in the
three-band model on a square lattice. Figure~(a) shows the lattice
structure, where each lattice site contains three orbitals and the
arrows represent
the breaking of the time-reversal symmetry. By putting the system on a torus and a cylinder, the single-particle energy spectra are shown in Figs.~(b) and (c). In Fig.~(c), 
chiral edges states  (thick lines) are observed.}\vskip-0.5cm
\label{fig:square}
\end{figure}

At $\Delta=0$, the time-reversal symmetry is preserved and two of the three bands cross
at the center (corner) of the Brillouin zone (BZ). For $\Delta>0$, the bands become gapped.
In order to ensure the ``flatness'' of
the top band, we require the energies are equal at the $\Gamma$ point, 
M point and X point,
which implies $\delta=-4 t_{dd}+2 t_{pp}+\Delta-2 t_{pp} \Delta/(4 t_{pp}+\Delta)$ and
$t_{pp}'=t_{pp}\Delta/(4 t_{pp}+\Delta)$. For simplicity, we set $t_{dd}=t_{pd}=t_{pp}=1$.
By varying $\Delta$, we found that the ratio of bandwidth/band gap
is minimized ($\simeq 1/20$) at $\Delta=2.8$.
Here the top and the bottom bands carry opposite Chern numbers $\pm1$
while the middle band has a trivial Chern number. The band structure
of this model is shown in Figs.~\ref{fig:square_band} and
\ref{fig:square_edge}, where the former is computed for periodic
boundary conditions (on a torus) and the latter on a
cylinder with two open edges. The edge states appearing in
Fig.~\ref{fig:square_edge} confirm the nontrivial Chern numbers
of the system.

\begin{figure}
\begin{center}
\subfigure[]{\includegraphics[width=0.15\textwidth]{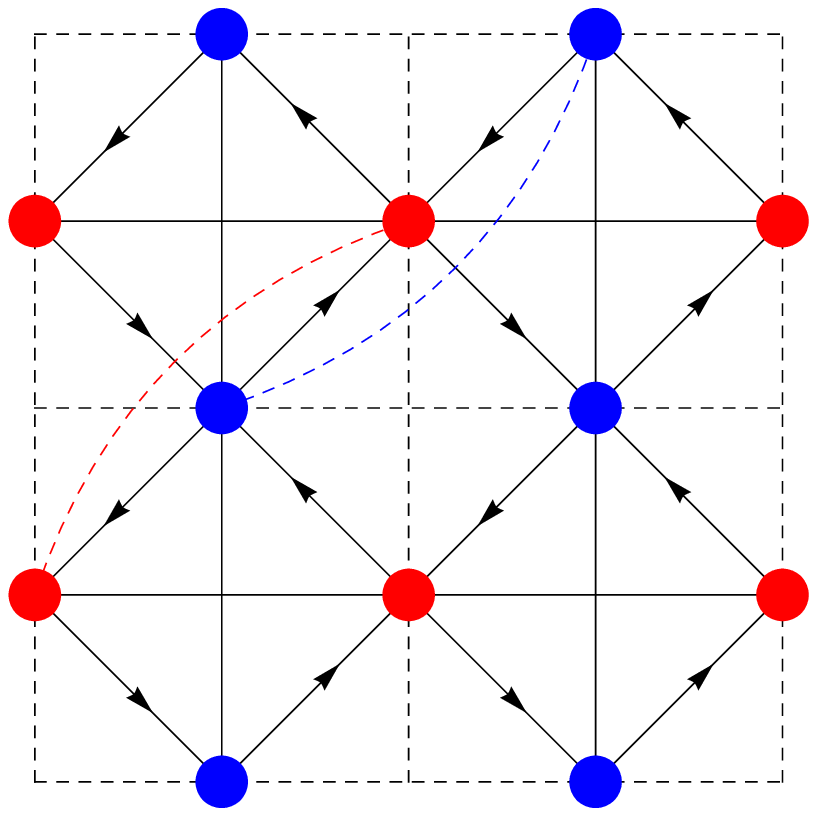}\label{fig:checker_board_lattice}}
\subfigure[]{\includegraphics[width=0.25\textwidth]{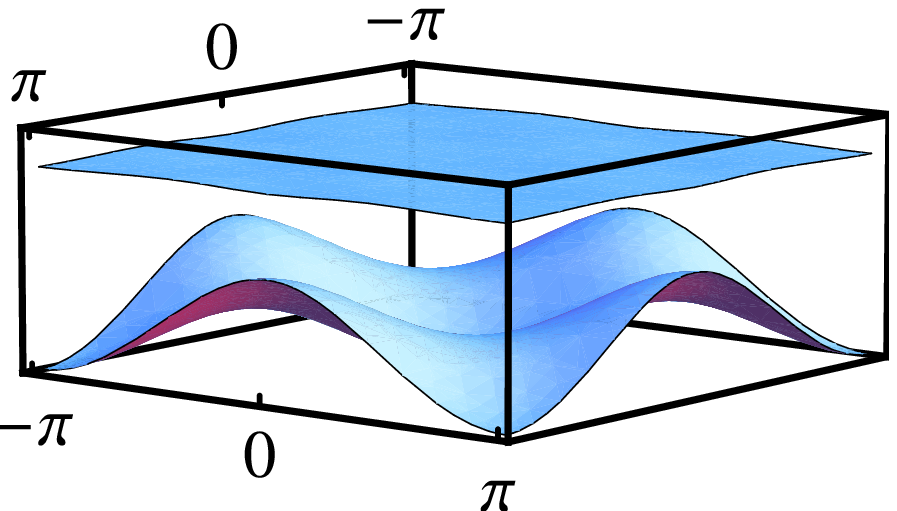}\label{fig:checker_board_band}}
\subfigure[]{\includegraphics[width=0.4\textwidth]
{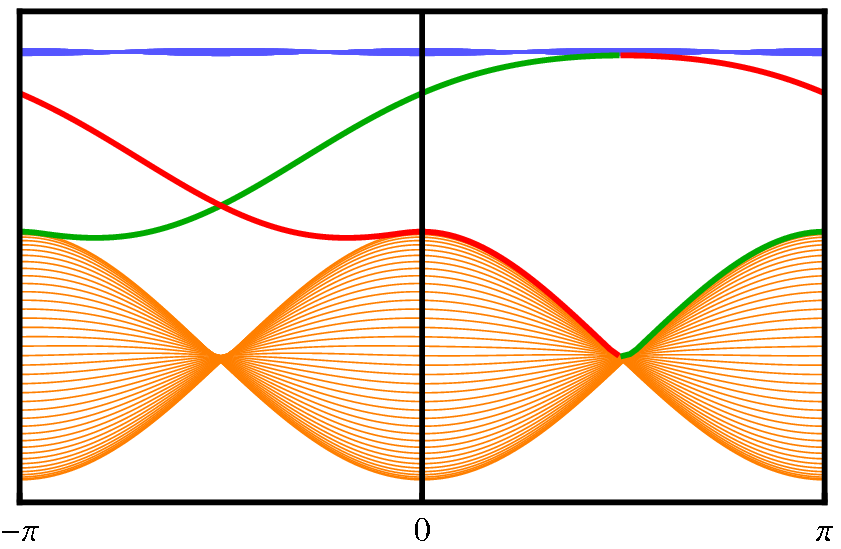}\label{fig:checker_board_edge}}
\end{center}
\vskip-0.5cm \caption{(Color online) Chiral quasi-flat band on the
checkerboard lattice. The lattice structure is shown in Fig.(a),
with arrows and (solid and dashed) lines representing the NN and NNN
hoppings respectively. The direction of the arrow shows the sign of
the phase in the NN hopping terms. Two of the NNNN hoppings are
shown as the dashed curve. Other convention are the same as in
Fig.~\ref{fig:square}.}\vskip-0.5cm
\end{figure}

{\it A two-band model on a checkerboard lattice}--- The model discussed above has three bands. Here we
present another model with only two bands. A two-band model has the
following advantages: (1) its band structure is much easier to
compute analytically; and (2) the Hilbert space is much smaller than
models with more bands and thus numerical studies become easier.
However, here we need to allow the next-nearest-neighbor (NNN) and
next-next-nearest-neighbor (NNNN) hoppings.  We emphasize
that single-band models can only have trivial Chern numbers and thus,
a two-band model is the minimal model to have topologically nontrivial
bands.

Consider a checkerboard lattice with NN ($t$),  NNN ($t'_{1}$, 
$t'_{2}$), and NNNN ($t''$) hoppings
[Fig.~\ref{fig:checker_board_lattice}]. Here, we allow the NN
hopping to carry nonzero complex phase ($\pm \phi$), whose
signs are shown
by the arrows in
Fig.~\ref{fig:checker_board_lattice}. These complex hoppings
break the time-reversal symmetry at $\phi\ne n \pi$ ($n \in \mathbb{Z}$).
The Hamiltonian of this model is
\begin{align}
H=&-t \sum_{\langle i,j \rangle} e^{i\phi_{ij}} (c^\dagger_i c_j+h.c.)
- \sum_{\langle\langle i,j\rangle\rangle}t'_{ij} (c^\dagger_i c_j+h.c.)
\nonumber\\
&- t'' \sum_{\langle\langle\langle i,j\rangle\rangle\rangle}(c^\dagger_i c_j+h.c.)
\end{align}
where $c_i$ ($c_i^\dagger$) is the fermion annihilation (creation)
operator at site $i$. The NN, NNN and NNNN sites are represented by
$\langle i,j \rangle$, $\langle\langle i,j \rangle\rangle$ and
$\langle\langle\langle i,j \rangle\rangle\rangle$. The phase factor
in the NN hopping terms is $\phi_{ij}=\pm \phi$ with the sign
determined by the direction of the arrows. The hopping
strength between NNN sites $t'_{ij}$ takes the value of $t'_1$
($t'_2$) if the two sites are connected by a solid (dashed) line.

The checkerboard lattice has two sublattices, and thus the Hamiltonian can be written in the momentum space as $H=-\sum_{\vec{k}} \psi_{\vec{k}}^\dagger \mathcal{H}\psi_{\vec{k}}$,
where $\psi_{\vec{k}}=(a_{\vec{k}}, b_{\vec{k}})$ is a two component spinor and $\mathcal{H}$
is a $2\times 2$ matrix
\begin{align}
\mathcal{H} &=[(t'_1+t'_2) (\cos k_x+\cos k_y) +4 t'' \cos k_x\cos
k_y] I
\nonumber\\
&+4 t \cos \phi (\cos\frac{k_x}{2} \cos\frac{k_y}{2})\sigma_x + 4
t \sin \phi (\sin\frac{k_x}{2} \sin\frac{k_y}{2})\sigma_y
\nonumber\\
&+(t'_1-t'_2) (\cos k_x-\cos k_y)  \sigma_z.
\end{align}
Here $I$ and $\sigma_{x,y,z}$ are the identity and Pauli
matrices.

At $\phi=0$ or $n \pi$, the time-reversal symmetry is preserved and
the two energy bands 
cross at the corner of the
BZ~\cite{sun2009}. At $\phi \ne n \pi$, a gap opens up
between these two bands. Due to the breaking of the time-reversal
symmetry, each band can carry a nonzero Chern number.
In order to reach a flat band, we require the energies of the top band are equal
at at the $\Gamma$ point, 
M point, K point and at $\vec{k}=(\pm \pi/2,\pm \pi/2)$.
With $t=1$ and $\phi=\pi/4$, this condition implies $t=1$, $t'_1=-t'_2=1/(2+\sqrt{2})$, $t''=1/(2+2
\sqrt{2})$. With these values,
the top band becomes very flat, with bandwidth of about
$1/30$ of the gap [Fig.~\ref{fig:checker_board_band}]  and each of the two bands
now carries Chern number $\pm 1$. We further verify this conclusion via the
study of the chiral edge mode [Fig. ~\ref{fig:checker_board_edge}].

{\it Discussion}--- In addition to the models discussed above,
similar effects can be observed in other models  with
quadratic touching. For example, if we allow NN and NNN
hoppings, both the kagome lattice and the honeycomb lattice with
the $p_x$ and $p_y$ orbtials~\cite{C_Wu,Zhao2008,Wu2008} can support this type of nearly-flat
bands when the time-reversal symmetry is broken.
Here the kagome-lattice model is a three-band one, while the other has
four bands.


Here, we compare the models with quadratic band touching and those with Dirac points.
Due to fermion doubling, the Dirac points need to appear in pairs.
In order to reach an insulating phase starting from a semi-metal with
two Dirac points, a nonzero mass need to be introduced at each of
the two Dirac points. However, depending on the relative sign of the
two Dirac masses, the resulting insulator can be either
topologically trivial or nontrivial~\cite{Haldane1988}. Due to this
uncertainty on the topological structure, the nearly-flat band from
a model with Dirac points may be topologically trivial and thus
irrelevant to our interests. On the contrary, for the models with quadratic band touching, 
the constraint of fermion doubling is absent and there is a 
single crossing point. This crossing point can be regarded 
as two Dirac points merging together, and is protected by
time reversal symmetry and discrete rotational symmetry, 
e.g., $C_4$ in the checkerboard lattice model. Since two 
hidden Dirac cones have the same chirality~\cite{sun2009}, the energy 
bands will have nontrivial topological numbers once the gap is 
opened by breaking time reversal symmetry by complex 
hoppings which do not break discrete rotational symmetry~\cite{note}.
Therefore, we can focus on the flatness of the band, without worrying about finding a
topologically-trivial band. This discussion is also the reason that
we considered the $d$-orbital in the three-band square lattice
model. Without the $d$-orbital, the two $p$-band have two quadratic
band crossings at the center and the corner of the BZ, instead of
just one. Therefore, the resulting insulator may be topologically
trivial.

\begin{figure}
\begin{center}
\includegraphics[width=0.46\textwidth]{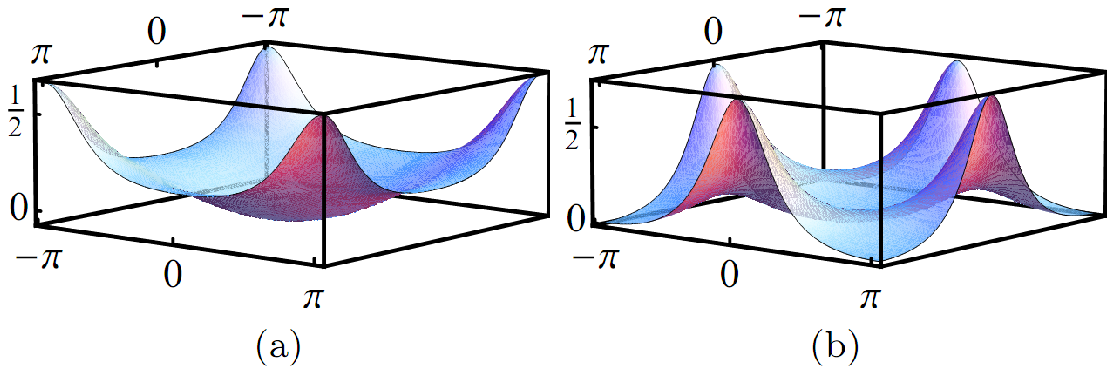}
\end{center}
\vskip-0.5cm \caption{ (Color online) Distributions of the Berry
curvature in momentum space for the flat bands
in (a) the square-lattice model and (b) the checkerboard lattice model. 
} \label{fig:Berry-curvature}\vskip-0.5cm
\end{figure}

Although the band gap is much lager than bandwidth in these models,
it is not clear whether such nearly-flat bands are equivalent to
Landau levels in 2DEG. However, the Berry curvature
(Fig.~\ref{fig:Berry-curvature}) in momentum space shows no sharp
features and the only length scale is the lattice constant, in sharp contrast
to the cases in which the Berry curvature has delta-function like peaks, 
e.g. Ref.~\cite{Onoda2002} . 
Thus, we argue that the topological nearly-flat bands we propose are very
similar to 2D Landau levels and we expect FQHE at fractional
fillings when repulsive interactions are turned on. We note in this
context that even 2D Landau levels have a short lattice length,
typically 10-100 times smaller than the magnetic length, underlying
the real physical 2DEG. A more detailed numerical study will be
presented in our future work.

When spin degrees of freedom are taken into account, the discussion
above can be generalized to the time-reversal invariant $\mathbb{Z}_2$
topological index by just substituting the time-reversal symmetry breaking terms into
corresponding spin-orbit couplings (where
spin up and down particles break the time-reversal symmetry in the
opposite way and thus the time-reversal symmetry is recovered when
both spin species are taken into account). In such a way, it may even be
possible to realize \emph{fractional topological insulators}~\cite{Levin2009} in these models.

{\it Experiment realization}---
In the discussion above, we only provided the optimum values for the parameters at which the flatness of the band is maximized. However, the nearly-flat band does not require strictly fine-tuning to reach. In fact, even if the parameters are changed by about $10\%$, the band remains to be fairly flat in the models we studied (see Supplementary Material for details). 
Because of this stability and the simpleness of these models, we believe that experimental realizations of
these models are possible in both condensed matter systems and ultra-cold atomic gases.
The insulating gap in these systems can be opened via spontaneous symmetry breaking if a
small amount of short-range repulsions are introduced~\cite{sun2009,sun2008,sun2010}. The same effect can  be expected via explicit symmetry breaking, e.g. by introducing a
magnetic field (for charged particles) or an artificial gauge field
(for charge neutral particles)~\cite{Stanescu2010}, as well as by rotating the lattice~\cite{Wu2008}.
In recent experiments, some optical lattices have been constructed whose band
structures are described by the 
square-lattice model and the 
honeycomb-lattice model discussed above~\cite{Wirth2010, Olschlager2010, Gemelke2010,Zhang2010}. Considering the fact that hopping strength can be tuned relatively easily in cold gases via varying the optical lattices, these cold-atom systems may be the leading candidates
for the realization of the topological physics predicted in our work.


{\it Acknowledgment}.---The authors would like to thank X-G. Wen and I.
Maruyama for valuable discussions. The work was supported by
JQI-NSF-PFC, AFOSR-MURI, ARO-DARPA-OLE, and ARO-MURI (K.S. and S.D.S.) and Z.C.G. is supported in
part by the NSF Grant No. NSFPHY05- 51164.

{\it Note added}.---
Related work has recently been done in Refs.~\cite{Tang,Neupert2010}. Very recently, the existence of fractional quantum Hall effect in our model has been confirmed by exact numerical studies as reported in Ref.~\cite{Sheng2011}.


\newpage
\onecolumngrid

\renewcommand{\thesection}{S-\arabic{section}}
\renewcommand{\theequation}{S\arabic{equation}}
\setcounter{equation}{0}  
\renewcommand{\thefigure}{S\arabic{figure}}
\setcounter{figure}{0}  

\setcounter{equation}{0}  
\setcounter{figure}{0}  

\centerline{\LARGE \bf{Supplementary Materials}}
\section{I. Topological flat band from long-range hopping}
If nonlocal terms are allowed, flat band with nonzero Chern number can be
constructed using an idea similar to the spectral flattening trick.
For an $n$-band model whose bands carry nontrivial Chern numbers,
e.g., Haldane's honeycomb lattice model, we can
write the Hamiltonian in momentum space as an $n \times n$ matrix
\begin{align}
\mathcal{H} ({\vec k}) = U ({\vec k}) \Lambda({\vec k}) U^\dagger
({\vec k}),
\end{align}
where $\Lambda({\vec k})$ is the diagonal matrix with
$E_a ({\vec k})$ ($a=1, ..., n$) on the diagonal and $U({\vec k})$
is a unitary matrix. From $\Lambda({\vec k})$, we can construct a
new diagonal matrix ${\tilde \Lambda} ({\vec k})$ in which the
energy of the topologically non-trivial band, say $E_b({\vec k})$,
is replaced with $E_b$, the constant independent of ${\vec k}$. Then
we construct a new Hamiltonian: $\tilde{\mathcal{H}}({\vec k}) =
U({\vec k}) {\tilde \Lambda} ({\vec k}) U^\dagger ({\vec k})$. It is
obvious that $\tilde{\mathcal{H}} ({\vec k})$ has a completely flat
band at $E= E_b$, which carries nonzero Chern number since the
eigenvector corresponding to $E_b$ remains unchanged from that of
$E_b ({\vec k})$. Note that as long as the $b$th band is isolated
from the other bands, the other energy levels in ${\tilde \Lambda}
({\vec k})$ can also be modified. One can obtain the Hamiltonian in
real space by the inverse Fourier transform of
$\tilde{\mathcal{H}}({\vec k})$. However,  in this construction, the
hopping amplitudes in general remain nonzero even between sites with
arbitrary large separation.

\section{II. Flatness of the band}

\begin{figure}[h]
\begin{center}
\includegraphics[width=0.8\textwidth]{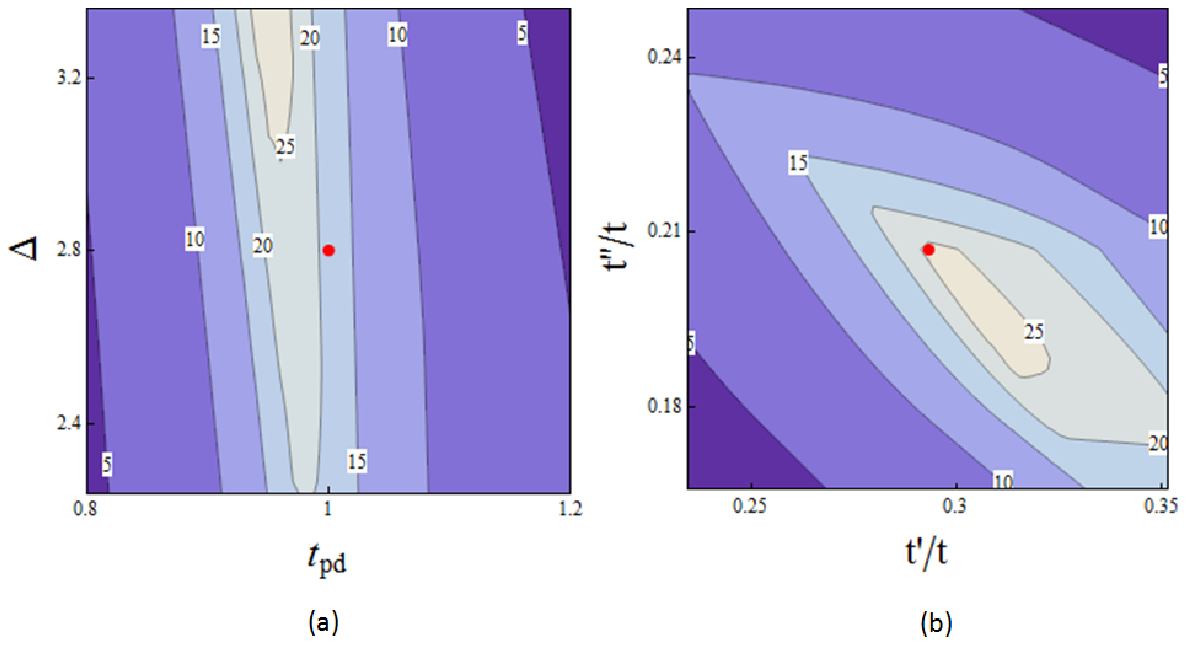}
\end{center}
\vskip-0.5cm \caption{(Color online) The ratio of band gap/bandwidth for
the nearly-flat band in (a) the square-lattice model and 
(b) the checker-board lattice model.  In Fig. (a), the horizontal and vertical axes are $t_{pd}$ 
and $\Delta$, correspondingly. For other parameters, we set 
 $\delta=-4 t_{dd}+2 t_{pp}+\Delta-2 t_{pp} \Delta/(4 t_{pp}+\Delta)$,
$t_{pp}'=t_{pp}/(4 t_{pp}+\Delta)$ and $t_{dd}=t_{pp}=1$. In Fig. (b), we set $t'_1=-t'_2=t'$ 
and $\phi=\pi/4$ with the two axes being $t'/t$ and $t''/t$. The dot at the center of each figure 
marks the parameters used in the Letter. 
In both figures, the parameters are varied by the amount of $\pm 20\%$. 
The ratio of band gap/bandwidth are marked in the figures for each contour.}\vskip-0.5cm
\label{fig:ratio}
\end{figure}

In Fig.~\ref{fig:ratio}, we show the contour plot of the ratio of band gap/bandwidth for 
the nearly-flat band at different parameters ($\pm 20\%$ away from the values we used 
in the Letter). As can be seen from the figures, in a wide region of the parameter space, 
this ratio remains large indicating a very flat band. 

We'd also like to emphasize that in this Letter, we do not intend to 
maximize the flatness of the band, but to demonstrate a generic technique which provides 
nearly-flat topological bands. For example, in the square-lattice model, we set 
$t_{dd}=t_{pd}=t_{pp}=1$ in the Letter for simplicity and maximized the ratio of 
band gap/bandwidth within this constrain by varying $\Delta$, which results in a very-flat band
with nontrivial topology. 
However, as shown in Fig.~\ref{fig:ratio}.(a), it is possible to enhance this ratio by removing 
the requirement of $t_{dd}=t_{pd}=t_{pp}=1$, 
but it is not the central focus of this work to find the optimum value for each control parameter
to maximize this ratio.
\end{document}